# The "I" in FAIR: Translating from Interoperability in Principle to Interoperation in Practice

Evan Morris[1], Gaurav Vaidya[1], Phil Owen[1], Jason Reilly[1], Karamarie Fecho[2], Patrick Wang[3], Yaphet Kebede[1], E. Kathleen Carter[1], Chris Bizon[1]

[1] Renaissance Computing Institute, University of North Carolina at Chapel Hill, North Carolina, USA 27599
[2] Copperline Professional Solutions, LLC, Pittsboro, North Carolina, USA 27312
[3] Department of Electrical and Computer Engineering, Duke University, Durham, North Carolina, USA 27708

## Abstract

The FAIR (Findable, Accessible, Interoperable, and Reusable) data principles [1] promote the interoperability of scientific data by encouraging the use of persistent identifiers, standardized vocabularies, and formal metadata structures. Many resources are created using vocabularies that are FAIR-compliant and well-annotated, yet the collective ecosystem of these resources often fails to interoperate effectively in practice. This continued challenge is mainly due to variation in identifier schemas and data models used in these resources. We have created two tools to bridge the chasm between interoperability in principle and interoperation in practice. **Babel** solves the problem of multiple identifier schemes by producing a curated set of identifier mappings to create cliques of equivalent identifiers that are exposed through high-performance APIs. **ORION** solves the problems of multiple data models by ingesting knowledge bases and transforming them into a common, community-managed data model. Here, we describe Babel and ORION and demonstrate their ability to support data interoperation. A library of fully interoperable knowledge bases created through the application of Babel and ORION is available for download and use at https://robokop.renci.org.

## 1 Introduction

Contemporary research is increasingly shaped by the availability of large, complex datasets. This is especially true in biomedical sciences, where advances in areas such as high-throughput technologies, electronic health records, population-scale biobanks, and molecular profiling produce an unprecedented volume and diversity of data. These resources offer extraordinary opportunities for discovery, enabling researchers to ask more ambitious questions and to approach longstanding problems with new analytic power. However, the scientific utility of these data is limited by the challenges of integration and reuse. In many biomedical domains, research findings are curated and presented as knowledge bases (KBs), which serve as structured repositories that represent biomedical entities, their relationships, and associated metadata in a formal, machine-readable framework. KBs generated in different contexts by different research groups, institutions, or consortia are often represented using distinct identifier systems, data models, and access mechanisms. Without careful harmonization, these differences impede integration, comparison, and the construction of unified representations of biomedical knowledge. As the number and diversity of publicly available biomedical knowledge bases continue to grow, the absence of scalable mechanisms for their integration increasingly constrains the scientific value of these resources.

The FAIR (Findable, Accessible, Interoperable, and Reusable) data principles [1] have emerged as a widely accepted framework for promoting the reuse of scientific KBs. FAIR encourages the use of persistent identifiers, standardized vocabularies, and formal metadata to enhance accessibility and interoperability. Although substantial progress has been made in findability and accessibility, practical challenges remain in the promotion of interoperability and reusability[2,3]. Achieving true interoperability requires not only the adoption of shared standards but also mechanisms that reconcile semantic and structural differences across knowledge bases. The absence of such mechanisms continues to constrain integration at scale.

Two key challenges in the integration and interoperation of knowledge bases are disparate identifier schemas and variation in data models. KBs often employ distinct identifiers or naming conventions to represent the same underlying biological concept, such as a gene, disease, or chemical. Effective integration requires that both users and computational systems recognize when different names or identifiers refer to the same entity. Beyond this, variations in data models, such as the types of information associated with each entity and the structures used to encode this information, introduce additional complexity. While identifier normalization has been widely recognized as a central challenge [4], aligning divergent data models presents an equally significant barrier to large-scale integration. Knowledge bases vary widely in their schemas, ontologies, and levels of granularity and these inconsistencies can obscure meaningful relationships and hinder the construction of a unified framework.

To address these challenges, knowledge base creators often design their systems to align with the FAIR principles for interoperability. These principles encourage, among other practices, that:

**I1**. (meta)data use a formal, accessible, shared, and broadly applicable language for knowledge representation.

**I2**. (meta)data use vocabularies that follow FAIR principles

**I3**. (meta)data include qualified references to other (meta)data

I2, for instance, may be satisfied by using gene identifiers from established public resources, such as EntrezGene [4] or Ensembl [5], which enable users to identify the genes in the knowledge base unambiguously. However, there is no universally accepted standard for identifiers. Moreover, different knowledge bases often divide the same conceptual space according to their specific priorities. One knowledge base may categorize water as an environmental feature, another as a pharmaceutical product, and yet another as a biochemical metabolite. Consequently, many knowledge bases are created using FAIR-compliant and well-annotated vocabularies, yet the collective ecosystem of these resources still fails to effectively interoperate in practice.

Extending interoperability in principle to interoperation in practice requires tools that can flexibly navigate between competing standards. In many cases, differences across knowledge bases reflect legitimate variation in priorities or underlying assumptions. However, other discrepancies are simply the result of independently developed models, introducing avoidable complexity that hinders integration without adding substantive value. Addressing this variability is essential to making large-scale, multi-source integration of knowledge bases truly feasible.

To this end, we have developed a suite of tools designed to *enable practical interoperation* across a range of knowledge bases that are individually FAIR-compliant [5,6] yet remain difficult to integrate collectively. These tools directly address the core barriers to data integration, namely, the need to (1) resolve entity equivalence and (2) harmonize differences in data models. Specifically, Babel and the related services Node Normalizer and Name Resolver address the first challenge through the robust normalization of identifiers. The Operational Routine for the Ingest and Output of Networks (ORION) addresses the second challenge through the transformation of diverse knowledge bases into *Knowledge Graphs* (KGs) using a community-maintained upper-level ontology and data model, Biolink Model [7]. Together, these components form a cohesive framework that automates processes which would otherwise require manual, source specific effort and lack generalizability at scale. This framework allows heterogeneous knowledge bases to be ingested, mapped to a unified system of identifiers, structured within a shared data model, and accessed through a consistent set of interfaces. By operationalizing the FAIR principles at each of these levels, our framework facilitates the construction of integrated, high-quality KGs that support rigorous scientific discovery.

# 2 Results

## 2.1 Babel: A Curated Framework for Identifier Equivalence

Babel is a set of hand-curated pipelines for downloading, filtering, and combining mappings from multiple sources across 23 broad classes of biological entities, ranging from genes and proteins to processes and pathways. Babel systematically processes equivalence information from a wide array of biomedical databases, ontologies, thesauri, and controlled vocabularies. It applies curated rules to synthesize these mappings into structured groups or "cliques", each representing identifiers believed to denote the same underlying concept. Within each clique, a single identifier is chosen as the "clique leader" or representative identifier, following a source preference order defined by Biolink Model [7] (see section "Identifier equivalence: Babel" under "Methods" below). Downstream sources can *normalize* identifiers by using the clique leader to represent this concept.

Given the scale and variability of biomedical vocabularies, curating each equivalence relationship individually would be infeasible; therefore, we prioritize assembling well-supported sets of equivalence assertions selected from trusted sources. Through a human-based curatorial process, decisions are made regarding which groupings of equivalences to accept from which mapping source, establishing a controlled framework for resolving identifier equivalences across domains. This process is expressly focused on identifier alignment and does not attempt to reconcile the full logical structures or relationship hierarchies embedded in source ontologies. This tradeoff reflects a conscious decision to prioritize scalable cross-cutting interoperability over exhaustive semantic granularity, which most source databases themselves do not maintain. By publishing cliques of equivalent identifiers as unified concepts rather than as a set of individual mappings, Babel enables practical interoperation by providing a shared vocabulary for downstream applications.

While perfect semantic equivalence is a valid goal, the practical integration of knowledge often requires a more pragmatic approach. In some cases, consumers of cliques require the merging of identifiers that other consumers prefer to keep distinct. We refer to this process as *conflation*: a controlled mechanism for grouping concepts that are closely related but not strictly identical. For example, the GeneProtein conflation provides cross-references for merging cliques representing protein-encoding genes with those representing the proteins they encode in order to simplify harmonization between gene and protein databases. This approach is particularly useful in analyses where distinguishing between a gene and its protein product offers little added insight while imposing extra costs to maintain and query a database, such as when investigating their combined role in a biochemical pathway. Similarly, the DrugChemical conflation merges cliques representing drugs with a single active ingredient with the chemical representing that active ingredient. The Babel pipeline generates conflations as lists of clique leaders to be treated as equivalent under defined conflation strategies.

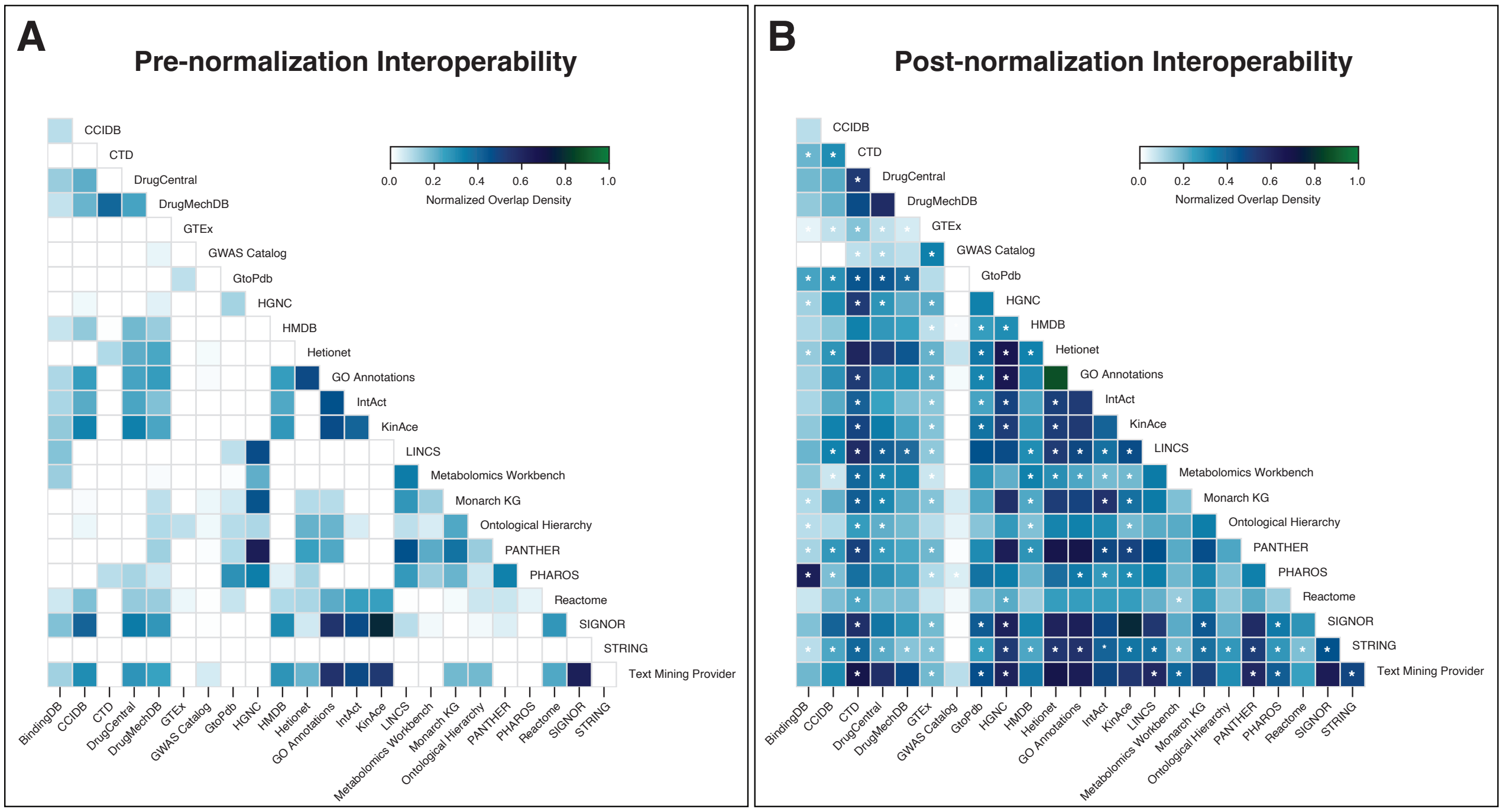

**Figure 1. Effect of Identifier Normalization on Cross-Source Interoperability** Knowledge base overlap heatmaps showing normalized co-occurrence density between source pairs (**A**) Pre-normalization overlap matrix with 138 connections. (**B**) Post-normalization overlap matrix with 267 connections. Each cell represents the scaled overlap density between sources A and B. Normalization resulted in 129 new source pair connections, representing a 93.5% increase in total overlap, defined as the number of unique source pairs that share at least one normalized entity. White asterisks (*) in panel **B** indicate new connections that were absent pre-normalization**.**

Downstream systems, including Node Normalizer (see "Applications of Babel for Identifier Resolution" below), can leverage this information to adjust or expand equivalence classes at query time depending on application requirements. This capability ensures that the framework supports both stringent and relaxed definitions of equivalence, thereby balancing analytic precision with practical large-scale data integration.

In addition to conflation, Babel is capable of enriching cliques with supplementary information that increases their analytic utility. The principal enrichment step assigns a single Biolink class to each clique. While this is straightforward for most pipelines, additional disambiguation is sometimes required, particularly for UMLS and chemical datasets, where equivalent identifiers may link entities from different conceptual categories, such as metabolites and general chemicals. Ubergraph calculates normalized information content values for each ontology class based on the number of subclasses and existential relations, scaled from 0 (for very general classes with many subclasses) to 100 (for highly specific classes with none). Babel incorporates these measures by assigning each clique the minimum value observed among its member identifiers, thereby providing a consistent indicator of semantic generality. Gene and protein cliques are further annotated with taxonomic information derived, respectively, from the NCBI Gene database and UniProtKB, ensuring that the species context is preserved during integration. These enrichments extend cliques beyond their identifier resolution, supporting analyses that depend on measures of specificity and biological context.

Babel's large-scale equivalence integration consolidates identifiers into structured cliques, enabling harmonized cross-resource referencing that underpins downstream KG construction and analytics. Across all fourteen pipelines, Babel constructed over 423 million distinct cliques, with clique leaders from 44 vocabularies (distinguished by CURIE prefix), including nearly 600 million unique CURIEs from 66 vocabularies (Supplemental Table 1).

The impact of Babel on cross-resource connectivity is shown in **Figure 1**. Pre- and post-normalization heatmaps display the density of shared identifiers between a set of commonly used knowledge bases, with normalization nearly doubling the number of sources with observed overlaps (from 138 to 267). To compare overlap strength across knowledge bases of different sizes, we defined normalized overlap density as:

$$density(A, B) = overlap_count(A, B)/\sqrt{|A| \cdot |B|} \qquad (1)$$

where $overlap_count(A, B)$ is the total number of shared identifiers between source $A$ and $B$, and $|A|$ and $|B|$ represent the respective unique identifier counts. The square-root scaling provides a geometric mean-based adjustment that controls for source size, yielding a size-independent measure of cross-source entity overlap. Normalization highlights how identifier harmonization through Babel substantially increases semantic connectivity across the knowledge bases, revealing new points of interoperability between previously isolated sources. As shown in **Figure 1 panel B**, white asterisks mark entirely new cross-source overlaps introduced following identifier normalization. In addition to the increase in unique overlaps, the average node degree rose from 11.5 to 22.25, indicating that each resource became connected to a greater number of distinct sources. Collectively, these results demonstrate that systematic equivalence resolution via Babel enhances interoperability across biomedical resources, both by establishing new connections and by strengthening existing ones.

### 2.2 Applications of Babel for Identifier Resolution

Babel's large-scale equivalence cliques are served via a pair of public services that address two fundamental needs in knowledge base integration: 1) normalizing identifiers to a well-defined clique of identifiers, and 2) resolving free-text strings to standardized identifiers with sufficient information to disambiguate among alternatives. While Babel's cliques can be ingested into general-purpose databases, the scale and complexity of biomedical data demand specialized infrastructure in knowledge base integration that can deliver these capabilities with precision and speed. To address this requirement, we developed two complementary services, Node Normalizer and Name Resolver, to expose Babel outputs as web services described using OpenAPI [8].

**Node Normalizer.** Node Normalizer provides an HTTP API that accepts a list of identifiers to normalize and returns the full cliques for those identifiers, including the clique leader, Biolink type, and additional properties such as the information content score. The service resolves heterogeneous identifiers to canonical clique leaders with optional conflation, eliminating normalization as a persistent bottleneck in large-scale workflows. By maintaining equivalence information in memory and exposing it through an API, Node Normalizer supports real-time normalization at scale. This functionality is critical for systems such as the NCATS Translator [9], where queries traverse multiple heterogeneous knowledge bases. An analysis of 19,043 queries submitted to a Node Normalizer instance over one week in June 2025 revealed that it took between 0.83 ms and 10.89 seconds to respond to queries, with an average response time of 352.33 ms per request. Each query in this dataset included up to 2,769 CURIEs for normalization, yielding an average of 5.68 ms per CURIE. A Jupyter Notebook showing examples of Node Normalizer usage is available in its GitHub repository.

**Name Resolver.** Name Resolver provides an HTTP API that accepts a text string and searches for cliques with either a preferred label or a synonym containing that text string. The service complements this functionality by enabling the resolution of textual inputs, mapping labels and synonyms to canonical identifiers through ranked search results. This allows the seamless integration of free-text terms with structured KG queries. For example, a query for "aspirin" returns a ranked set of concepts that may include CHEBI:15365, PUBCHEM.COMPOUND:2244, and related drug formulations, with sufficient contextual information to distinguish among them. Similarly, a search for "ADA" can distinguish between the gene *adenosine deaminase* (NCBIGene:100), the disease adenosine deaminase deficiency (MONDO:0011130), and the orchid genus *Ada Lindl* (NCBITaxon:125078, now classified under *Brassia*). By providing contextual metadata such as Biolink type, preferred labels, and synonyms, Name Resolver enables both automated systems and human users to disambiguate terms and map them to computable identifiers. A Jupyter Notebook showing examples of Name Resolver usage is available in its GitHub repository.

Together, Node Normalizer and Name Resolver make Babel's harmonized equivalence sets actionable across both automated pipelines and user-facing exploration environments. By grounding identifier resolution in Babel cliques, these services translate the harmonization of identifiers into operational infrastructure, enabling true interoperability among biomedical data resources in practice.

### 2.3 Model Equivalence: ORION

Even if knowledge bases utilize shared identifier namespaces, they rarely conform to the same semantic models or data schemas. Relationships between biological entities are described using a diverse range of terminologies and ontologies, which vary in their level of granularity and specificity. Data is made available in a variety of formats that are often incompatible with each other, such as through non-standardized web-based APIs or files encoded in JSON, XML, or CSV formats. Furthermore, many important sources of biological knowledge provide data in formats that lack formal data models entirely, such as results from statistical analyses.

Operational Routine for the Ingest and Output of Networks (ORION) is a KG pipeline that addresses these challenges by transforming disparate biological data representations into modular **interoperating** KGs using a shared semantic framework, Biolink Model.

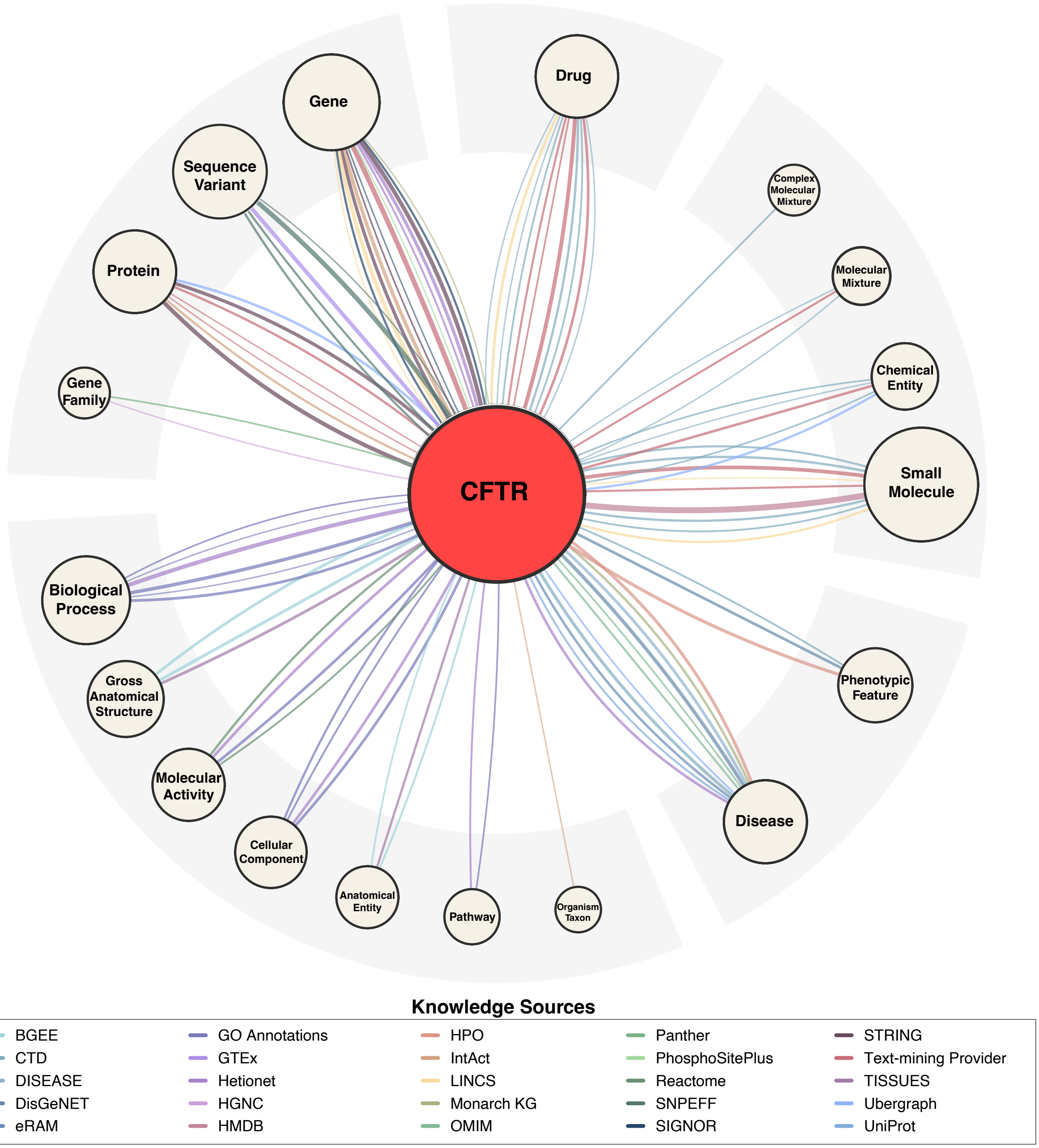

**Figure 2. Network representation of CFTR-linked components in the ROBOKOP KG generated by ORION.** In this example, knowledge from 25 of the sources integrated by ORION connect *CFTR* to 18 distinct biological node types through 35 unique relationship types. All entities and edges are modeled using the Biolink Model schema and normalized with Node Normalizer, enabling seamless exploration across knowledge bases. *CFTR* is shown as the central node, with surrounding nodes representing distinct biological entity types retrieved in response to a single database query. Nodes are grouped into higher-level categories, including chemicals, drugs, genes, proteins, biological contexts, diseases, and phenotypes. In the visualization here, node size reflects the number of entities of that type connected to CFTR, using a log-transformed count. Edges link CFTR to each entity type and are colored by the relationship source. Edge width indicates the number of edges between CFTR and each source–component pair, also log-transformed.

The ORION pipeline has been applied to over 40 public knowledge bases. It converts relationships expressed in heterogeneous data schemas to semantic triples, i.e., subject-predicate-object relationships, strictly defined by Biolink Model. ORION utilizes Node Normalizer to standardize identifiers used by knowledge bases into a shared, consistent identifier namespace and assign Biolink Model categories such as Gene or Disease to each entity. As a result, KGs produced using ORION are fully interoperating because they are consistent in both identifier usage and semantic structure and can be easily integrated or queried using a consistent and predictable model, thereby supporting a range of downstream applications. These modular KGs can be utilized independently or combined using ORION to construct integrated networks that retain source-level distinctions while enabling seamless interoperability.

To make KGs immediately usable, ORION generates output formats that are ready for deployment. Each KG is serialized in the KGX format [10] and accompanied by metadata files that capture versioning, provenance, summary statistics, and specific information about the transformations that were performed. In addition to KGX files, ORION optionally generates Neo4j database dumps, which can be imported and queried as standalone Neo4j instances. ORION-generated KGs can be deployed using model-aware KG querying tools such as Plater [11], which is a web service we developed to expose Biolink Model KGs through an API supporting a variety of query types, including TRAPI [12], Cypher, and custom endpoints for extracting information from KGs. Due to the usage of a consistent data model, tools such as Plater can apply uniform processes and APIs to deploy and expose the ORION-generated KGs. Furthermore, the adoption of a standardized data model enables reusable analytic pipelines without requiring additional transformations or reformatting.

One example application of ORION is the Reasoning Over Biomedical Objects linked in Knowledge Oriented Pathways (ROBOKOP) KG, a large-scale biomedical KG constructed entirely from ORION-produced components. The ROBOKOP KG contains approximately 10 million nodes and 130 million edges, aggregated from over 40 biological data sources. Because every source included in the ROBOKOP KG has been semantically aligned and identifier-normalized, the full graph can be utilized without requiring custom harmonization, supporting applications ranging from chemical-gene-disease analysis to semantic subgraph exploration across biological scales. The graph is available for download in KGX format and as a Neo4j database dump, accompanied by metadata. These resources enable researchers to utilize the ROBOKOP KG directly or as a foundation for creating new fit-for-purpose graphs.

To demonstrate how interoperation supports discovery, consider a researcher exploring a single gene of interest, such as *CFTR*, to examine its relationships with phenotypes, chemicals, molecular pathways, and other biological features. Using ROBOKOP KG, the researcher can retrieve all relevant relationships in a single database query. As shown in **Figure 2**, this returns connections from *CFTR* to 18 distinct types of biological entities, drawn from 25 different knowledge bases.

In a second use case, a researcher aims to assemble a dataset of relationships linking chemicals to diseases or phenotypes both directly and through their molecular targets. This task may involve integrating chemical-disease associations from the Comparative Toxicogenomics Database (CTD) [13], chemical-target interactions from BindingDB [14], and gene-phenotype relations from the Monarch Initiative[15]. Although each of these sources is publicly available and nominally interoperable, their practical integration presents significant challenges. CTD represents chemicals with MeSH identifiers and diseases with MeSH or OMIM terms, using free-text annotations to denote relationship types. BindingDB utilizes InChIKeys or other chemical identifiers and reports protein targets as UniProt accession numbers, with relationships inferred implicitly from the data structure. Monarch Initiative, in contrast, provides semantically explicit Biolink Model–compliant gene-phenotype associations using HGNC and HPO identifiers.

Even for an experienced informatician, merging these datasets generally requires familiarity with each source's data schema, utilization of external concordances, and programming custom scripts. For researchers without computational expertise, these obstacles can be prohibitive. In contrast, ORION enables users to specify the desired knowledge bases and automatically generate a coherent graph with consistent identifiers and a shared data model through a single command in Python or from the command-line interface. The resulting unified graph can be loaded into a database, visualized, or exported into other formats such as RDF using LinkML tooling.

The effect of this integration is illustrated in **Figure 3**. The left panels show the source-specific networks from BindingDB, CTD, and the Monarch Initiative, each of which captures a distinct subset of relationships among core Biolink entity types such as genes, proteins, chemicals, and diseases. The right panel displays the integrated graph produced by ORION, where entity identifiers have been normalized using Node Normalizer and relationships semantically aligned to Biolink Model. The merged network reveals new cross-source connections and shared nodes, demonstrating how harmonization across schemas transforms previously isolated knowledge bases into a coherent, interoperable structure that supports unified exploration and analysis.

By aligning identifiers, harmonizing relationship semantics, and producing modular, interoperable artifacts, ORION lowers the technical barriers to integration and secondary use across the biomedical ecosystem. ORION-generated KGs adhere to Biolink Model and utilize normalized identifiers, ensuring structural and semantic consistency across sources. This uniformity enables users to focus on asking biologically meaningful questions and conducting deep exploration of interdisciplinary knowledge, rather than resolving incompatibilities in knowledge bases. ORION thus advances the broader goal of making biomedical knowledge truly FAIR by transforming heterogeneous data into standardized interoperable resources that promote transparent and efficient discovery.

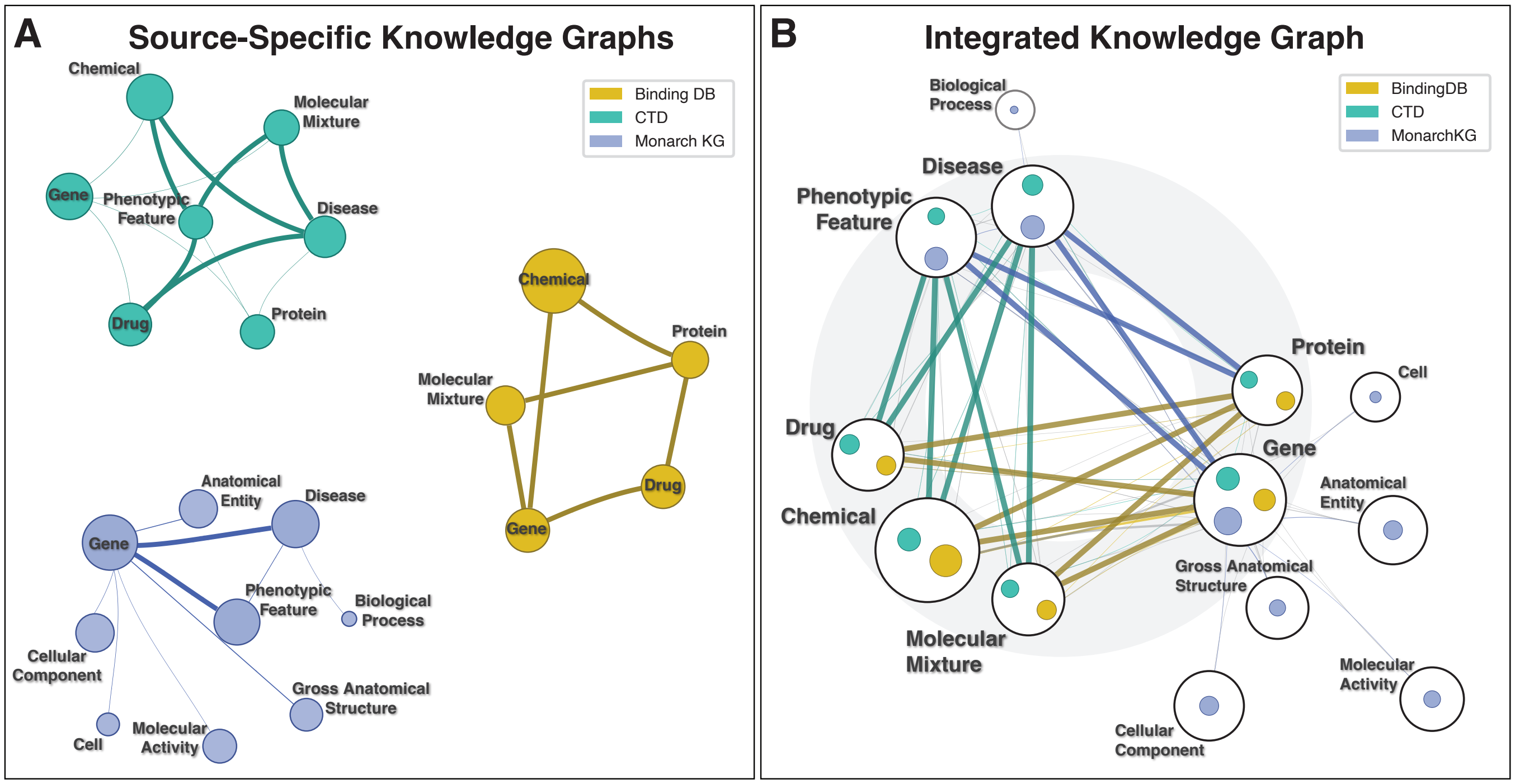

**Figure 3. Comparison of source-specific and integrated knowledge graphs generated by ORION.** (**A**) Network representations illustrate how three biological knowledge bases: BindingDB (gold), CTD (teal), and Monarch KG (blue) connect key Biolink entity types, including genes, proteins, diseases, and chemicals. The left panels show the intra-source networks derived from each resource, highlighting differences in coverage and relational structure. (**B**) The right panel displays the integrated graph generated by ORION, where entities and relationships have been normalized and semantically aligned using Biolink Model. Shared nodes demonstrate interoperability across sources, while edge density reflects newly created cross-source connections resulting from harmonization.

## 3 Methods

### 3.1 Identifier Equivalence: Babel

A foundational challenge in the integration of biomedical data sources lies in the resolution of entity equivalence. The same underlying biological concept may be represented by different knowledge bases using distinct identifiers, reflecting their specific priorities and use cases. For example, integration across heterogeneous sources requires understanding that the Chemical Entities of Biological Interest (ChEBI) [16] identifier “15377” and the PubChem identifier “962” both refer to the same concept - water. Within this paper, identifiers are expressed as Compact Uniform Resource Identifiers (CURIEs), for example, CHEBI:15377 and PUBCHEM.COMPOUND:962.

Existing resources often provide partial solutions. To facilitate integration, many databases and ontologies include cross-references that link their identifiers to those used by other resources. For instance, the ChEBI record for water includes cross-references to the concept of water as a chemical in many other databases, including PubChem [17] (PUBCHEM.COMPOUND:962), the Chemical Abstracts Service (CAS) Registry [18] (CAS:7732-18-5), the Human Metabolome Database [19] (HMDB:HMDB0002111), and the Kyoto Encyclopedia of Genes and Genomes [20] (KEGG.COMPOUND:C00001). However, it notably omits UNII and RxNorm identifiers, which are used to identify drug ingredients and formulations. Mappings may be synthesized by integrative resources such as the Unified Medical Language System (UMLS) [21], TogoID [22,23], or the EMBL-EBI Ontology Xref service (OxO) [24], or maintained externally through efforts like Mapping Commons [25], which use the Simple Standard for Sharing Ontology Mappings [26] to publish mappings between existing resources. Nonetheless, and are occasionally incompatible, creating gaps and conflicts that underscore the limitations of relying on any single resource alone.

Many biomedical concepts also exist within rich hierarchies or part-whole structures. For example, the National Library of Medicine (NLM) Medical Subject Headings (MeSH) thesaurus [27] distinguishes a general concept for water (MeSH:D014867) from more specific subconcepts such as carbonated water (MeSH:D061545), drinking water (MeSH:D060766), ice (MeSH:D007053) and steam

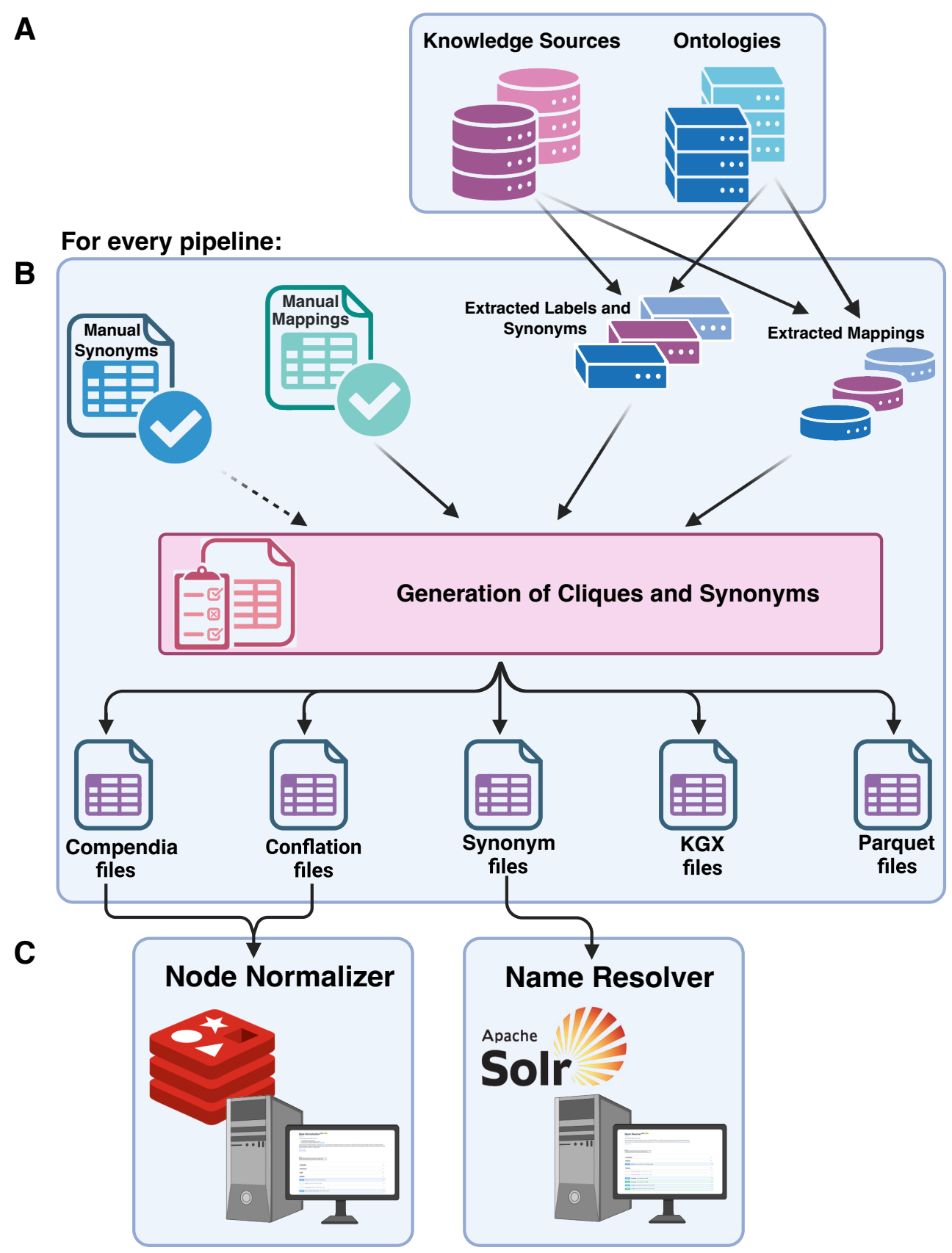

**Figure 4. Babel.** The overall data flow within Babel. (**A**) Labels, synonyms, and mappings are extracted from knowledge bases and ontologies. (**B**) Each of the eighteen output-specific pipelines takes some subset of these mappings, combines them with manual mappings and synonyms, and generates cliques of equivalent identifiers, along with all the synonyms associated with those identifiers. The cliques are used to produce compendia and conflation files (used by the Node Normalizer frontend) as well as synonym files (used by the Name Resolver frontend) (**C**). It also produces cliques in additional output formats (KGX and Apache Parquet).

Comprehensive integration that respects these distinctions would require explicit encoding of *subclass* or *part-of* relationships to clarify the scope of each identifier. While graph-based projects such as Ubergraph [28] have explored incorporating these complexities, most biomedical resources do not provide this level of semantic precision. Furthermore, vocabularies frequently differ on whether two identifiers should be considered equivalent at all, reflecting diverse domain perspectives.

Babel is a Python-based pipeline developed to address the pervasive challenge of identifier equivalence in biomedical data integration, where disparate resources frequently employ distinct identifiers or naming conventions to represent the same biological concept. The system is implemented using Snakemake [29], which structures the workflow as a directed acyclic graph of interdependent rules to automate the retrieval, processing, and synthesis of equivalence information from a diverse range of biomedical databases, ontological thesauri, and controlled vocabularies (see sources in Supplemental Table 1). Rather than attempting to reconcile the broader

(MeSH:D013227), while other resources only use a single concept for every form of water (e.g. CHEBI:15377).

logical hierarchies embedded in source ontologies, Babel focuses explicitly on constructing sets of identifiers, referred to as cliques, that represent strict equivalence. This design choice avoids contradictions that may arise from naive ontology merging and provides a consistent foundation for downstream applications that require harmonized identifiers.

Babel is structured as a single Snakemake workflow comprising 18 output-specific workflows, each represented by its own Snakemake file (**Figure 4**). Most of these workflows correspond to individual Babel pipelines, which are responsible for generating cliques in several closely related biomedical categories organized by Biolink type, such as the Anatomy pipeline, which produces cliques relating to four Biolink classes: AnatomicalEntity, GrossAnatomicalEntity, CellularComponent and Cell (see complete list in Supplemental Table 2). Two workflows generate the optional GeneProtein and DrugChemical conflations as sets of preferred identifiers that are combined when that conflation is activated (as described in the "Results" section previously). Five auxiliary workflows are not tied to any particular Biolink type but rather provide shared rules for downloading and transforming biomedical resources, generating reports or exports, and other shared steps. This modular layout offers several benefits: it allows individual Biolink types to be generated or debugged in isolation from others; it enables unrelated rules to run in parallel; and it supports modules where complex logic is required to choose a Biolink type for each clique such as for chemicals, where the distinction between molecular mixtures, small molecules and other chemical entities can require examining the chemical formula or comparing opinions from multiple sources. As a final step, all cliques and synonyms are loaded into a DuckDB [30] database so that duplications or identical labels can be noted.

Each Babel pipeline starts with a dedicated set of Snakemake rules that download and extract relevant content into standardized intermediate representations. For example, for chemical entities, the ChEBI pipeline orchestrates the retrieval of complete dataset exports from the ChEBI database [16], including structural files and accession tables, followed by parsing to extract labels (the primary textual name of a concept), synonyms (secondary textual names), and other metadata (e.g., descriptions, secondary identifiers, etc.). Domain-specific pipelines subsequently process these downloads to isolate identifiers, equivalence mappings, and synonym information that serve as inputs to the equivalence construction stage.

The extraction of mappings does not include every mapping from a source. That approach would lead to a compounding of errors and a set of very large cliques that contain unrelated items. Instead, human curators select a minimal set of mappings from each source that enable each identifier from the source to be associated with an identifier from another source. Particularly well-curated sources such as MONDO [31] or UniChem [32] may serve as a mapping hub, with human curators choosing to ingest many types of mappings from the source. The mapping curation is not performed at the level of individual identifiers, which would be impractical at scale. Instead, curators make higher-level decisions between entire vocabularies, such as whether Babel should accept MONDO's mappings from MONDO to Disease Ontology [33]. Supplemental Table 3 summarizes the breadth of cross-references integrated into Babel, along with the filtering strategy employed and the number of cross-references extracted from each. New identifier sources can be requested by creating a GitHub issue in the Babel GitHub repository, which also includes other guidance on contributing to Babel.

Once mappings have been extracted, Babel synthesizes the information into structured equivalence cliques. Each clique represents a group of identifiers determined to refer to the same biomedical concept. Within each clique, identifiers are filtered and prioritized according to a curated hierarchy defined by Biolink Model [7], which specifies valid prefixes for each Biolink class as implemented in Biolink Model Toolkit [34]. One identifier, represented as a CURIE, is designated as the preferred identifier for the clique, selected according to Biolink Model's preferred prefix order. Likewise, one label is chosen as the preferred label based primarily on the preferred prefix order specified for the Biolink type of the clique, but modified by several quality checks (such as known sources of good labels and suppression of long labels). This approach ensures stable and standardized identifier resolution and clear labels across diverse knowledge bases while intentionally avoiding broader logical entailments beyond strict equivalence.

The two conflation files are generated by using specific mappings to connect genes with proteins (the ID Mapping download from UniProt [35], which contains UniProtKB identifiers to NCBIGene identifiers) and drugs with chemicals (using RXCUI-based mappings from RxNorm [36] and UMLS). For GeneProtein conflation, genes are always listed before proteins. For DrugChemical conflation, the algorithm arranges the identifiers in a custom Biolink type order that prefers more specific types (e.g., Small Molecule) over more general types (e.g., Chemical Entity). Within each Biolink type, identifiers are sorted by information content, clique size, and CURIE numerical suffix.

Upon completion of clique and conflation construction, Babel produces multiple categories of output. Clique compendia are written as newline-delimited JSON files [37], each representing a single clique and its ordered list of equivalent identifiers. Corresponding synonym files record the preferred identifier for each clique alongside all recognized textual synonyms, enabling robust name-based resolution. Conflation outputs enumerate sets of clique leaders to be treated as equivalent under specific, explicitly defined conflation policies. All outputs are loaded into DuckDB [30] and exported as Apache Parquet files [38], which enable index-wide checks, such as identifying identifiers that appear in multiple cliques or cliques sharing identical preferred names, which are written out as machine-readable reports. Additionally, Babel produces KGX-formatted [10,39] exports that represent equivalence relationships as explicit node-and-edge structures aligned with knowledge graph conventions.

### 3.2 Applications of Babel for Identifier Resolution

**Node Normalizer.** The cliques generated by Babel can be directly ingested into any document-based database, such as ElasticSearch or MongoDB, for querying. However, because identifier normalization can be a critical bottleneck for query-time integration, we built a dedicated high-speed system to ensure rapid normalization.

Node Normalizer is a high-performance system that combines an in-memory Redis database and an OpenAPI-compliant FastAPI [40] application for identifier normalization. Python scripts load the Babel cliques, in JSON format, into Redis. The stored content includes most Babel clique data except for synonym information. Seven Redis tables are created: (1) a mapping from identifiers to their individual labels; (2) a mapping from identifiers to their normalized identifiers (the clique leaders); (3) a mapping from clique leaders to their Biolink types; (4) a mapping from clique leaders to additional properties such as descriptions, information content values, and taxa; (5) a table containing metadata about the database, including the distribution of prefixes for each Biolink type; and (6–7) two tables containing lists of identifiers that should be conflated under the GeneProtein and DrugChemical conflation strategies.

Users of Node Normalizer can query it with any number of identifiers to return their Biolink type, a list of equivalent identifiers, a preferred identifier and label and optionally the information content of the clique and the taxa in which it is found. Either or both GeneProtein and DrugChemical conflation may be turned on when running these queries. Additional API endpoints provide information about the data in the Node Normalizer database, such as which version of Babel outputs it contains, which conflations are supported and which Biolink types and CURIE prefixes are included. A Jupyter Notebook showing examples of Node Normalizer usage is available in its GitHub repository.

**Name Resolver.** Users of Name Resolver can query it with a text string in order to return a list of possible matches from the database, searching both the preferred name for each clique as well as all known synonyms. This allows it to be used to find biomedical identifiers corresponding to particular biomedical terms, either in a user-controlled manner (such as by filling in autocomplete suggestions for a user entering biomedical terms for a query) or in a bulk manner (such as to perform entity linking for a list of biomedical terms). Various parameters allow the results to be filtered by Biolink type, CURIE prefix or to a particular set of taxa. Highlighting and debugging options allow additional information about the match to be returned as well. A second API endpoint can return all known synonyms for a particular preferred CURIE. A Jupyter Notebook showing examples of Name Resolver usage is available in its GitHub repository.

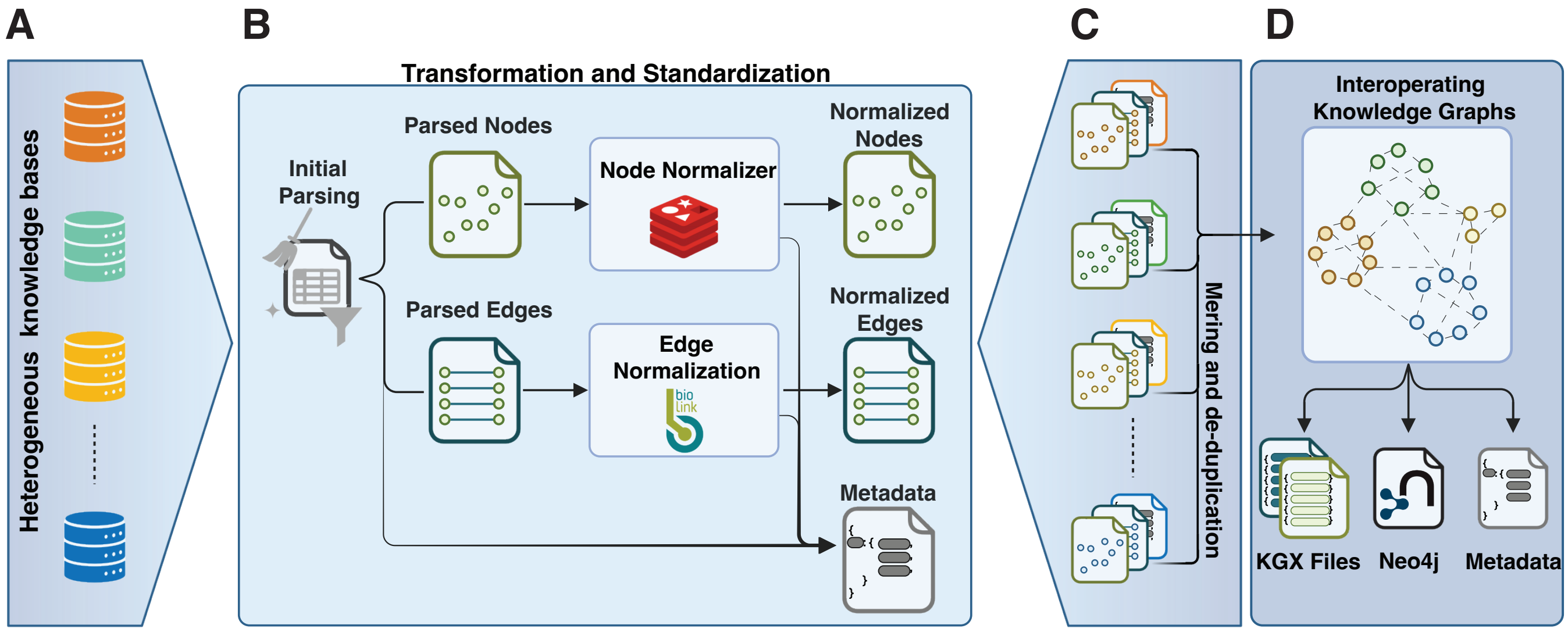

**Figure 5. ORION pipeline.** ORION ingests heterogeneous source knowledge bases (**A**) using source-specific parsers that retrieve versioned data releases and transform them into a shared intermediate representation (**B**). These parsed node and edge files are not required to be fully compliant with the Biolink Model but must satisfy minimal constraints to enable uniform downstream processing. Source-independent normalization aligns identifiers and semantics across all inputs, resolving node equivalence via Node Normalizer or the ClinGen Allele Registry and harmonizing predicates using the Biolink Model Toolkit. Normalized graph artifacts are serialized in the KGX JSON Lines format, while metadata capturing versioning, provenance, and quality metrics are generated throughout the pipeline. Normalized outputs are integrated across sources through union and de-duplication to produce an interoperable knowledge graph (**C**), which supports multiple downstream data products and access modalities (**D**).

Name Resolver provides an Apache Solr–based search engine for the labels and synonyms contained within Babel cliques. The service includes an OpenAPI-compliant FastAPI application written in Python that supports both individual and bulk queries, as well as synonym lookups in which all labels and synonyms for a preferred identifier are returned. A lightweight ingestion module prepares the Apache Solr database by loading the Babel clique data in JSON format. To reduce database size, only the preferred identifier, Biolink type, taxa information, and synonyms are included in this index. Equivalent identifiers, descriptions, and information content values are retrieved by subsequent lookup in Node Normalizer instead.

The Solr backend uses Apache Lucene to index labels and synonyms in two parallel ways. It uses the standard Solr tokenizer to break each synonym into individual words, while also employing a KeywordTokenizer to index each synonym as a single token, thereby supporting case-insensitive exact matches. Queries are executed using the eDisMax parser. When operating in autocomplete mode, the parser also identifies matches where the input string is incomplete. Boosting parameters prioritize exact matches over keyword matches, and preferred labels over synonyms.

### 3.3 Model Equivalence: ORION

ORION is a Python-based package for building and manipulating interoperable knowledge graphs (KGs). The core functionality of ORION is a pipeline for transforming data from original knowledge bases into node-normalized Biolink Model–compliant KGs (**Figure 5**). Because knowledge bases rarely conform to a standard schema, each must be handled using a custom module of code and configuration tailored to its specific format and semantics. For every knowledge base ingested by ORION, a dedicated parser retrieves a versioned data release, reads the original structure, and transforms it into a standardized knowledge graph representation. These preliminary representations are not required to be fully compliant with Biolink Model or use standardized identifiers. However, node identifiers must be resolvable by Node Normalizer or the ClinGen Allele Registry [41], and edge predicates must either be compatible with Biolink Model or mappable to it using available tooling [34]. This minimal set of constraints, along with a shared file format, ensures that all downstream normalization and transformation steps can be carried out uniformly across all sources.  A list of current source parsers is available in Supplemental Table 4 and in the ORION GitHub repository  [42].

After initial parsing, ORION applies a set of source-independent transformations to normalize identifiers and align semantics. Most entity identifiers are standardized using the Node Normalizer described above, which resolves each input to a preferred CURIE, a canonical label, synonyms, and an assigned set of Biolink Model categories. Sequence variants, which fall outside the scope of Node Normalizer, are processed via the ClinGen Allele Registry [41] and are assigned ClinGen Canonical Allele IDs as node identifiers. Predicates are harmonized using the Biolink Model Toolkit [34], which maps them to Biolink Model–compliant representations and attaches additional qualifiers such as direction, aspect, or context when applicable. The resulting output is a normalized, Biolink-aligned KG that preserves source-specific content while enabling cross-source interoperability. Graphs are serialized in the KGX JSON Lines format [10,39], which can be utilized by a range of downstream applications.

Because each ORION graph adheres to a consistent identifier scheme and a shared semantic model, integrating multiple knowledge bases is straightforward. Users provide an input specification yaml file, called a *graphspec*, which defines the set of knowledge bases to include. ORION constructs the integrated graph by taking the union of the nodes and edges from the specified component graphs. During this process, an edge-merging strategy is applied in which two edges are considered equivalent if they have the same subject, object, predicate, qualifiers, and primary knowledge base. This strategy unifies repeated assertions while preserving independent contributions from different sources.

A *graphspec* can also be used to specify other aspects of graphs and how they are built in a declarative way. Users can optionally provide specific versions of source data, Biolink Model, or normalization configuration parameters for each graph. Normalization configurations can be used for various identifier conflation and fallback behaviors (see above for details). If versions are omitted, ORION automatically detects and uses the latest available version of each knowledge base and Biolink Model.

The *graphspec* also allows users to configure edge-merging strategies. For example, a user may choose to include only those relationships from a given knowledge base that connect to entities found in other sources. This selective approach is employed in ROBOKOP KG [43], which incorporates subclass relationships from Ubergraph [28] but restricts inclusion to those that pertain to entities present in the integrated graph. This avoids unnecessary expansion while preserving relevant hierarchical information. The *graphspec* also enables property propagation across sources. For example, in ROBOKOP KG, subclass annotations from MONDO [31] and role designations from ChEBI [16] are transformed into node properties and applied to all nodes that share a synonymous identifier, thereby enriching the graph with curated metadata without requiring full source duplication.

Metadata are systematically generated and consumed throughout the ORION pipeline to support provenance tracking and quality control. Each graph artifact is versioned, and these version identifiers are recorded in the metadata files. ORION also captures process metadata, including the versions of the tools used during transformation, as well as quality metrics that summarize key characteristics of the resulting graph. These include counts of nodes and edges by type, as well as the proportion of entities that were successfully normalized. Tracking these metrics across successive builds allows users to detect unexpected deviations, such as a decline in normalization success or a sudden increase in node count, which may indicate schema changes in a source or issues in data acquisition. Finally, each ORION-produced KG is accompanied by a distributable metadata file containing provenance, versioning, and licensing information, to ensure adherence to FAIR principles.

Using the metadata and versions associated with each pipeline artifact, ORION implements a custom asset dependency management system. This ensures that the pipeline does not duplicate processes that were already performed, such as downloading or parsing source data or performing normalization. When new versions of data or Babel become available, ORION is able to programmatically determine which stages of the pipeline need to be run and only perform the necessary updates.

Because all ORION outputs adhere to a shared data model, the system also provides a foundation for downstream model-aware graph transformations. In particular, ORION supports the materialization of logically entailed relationships using Biolink Model's hierarchy. For example, when a downstream application benefits from a fully redundant graph in which all subclass relationships are explicitly included, these edges can be instantiated through tooling provided by the Biolink Model Toolkit [7]. This capability enables users to tailor the structure of the integrated graph to meet the reasoning requirements of their specific use case, whether that involves working with compact representations or expanded graphs enriched with inferred relationships.

ORION was designed with compute resource management in mind and can generally be used with low resource requirements (e.g., 16GiB of memory). Processing of large data sources and generic pipeline components, such as identifier normalization, is implemented using streaming techniques, which hold a minimal amount of data in memory at a time. ORION utilizes the JSON Lines file format for storing knowledge graphs to facilitate streaming. In some cases, where streaming data is not possible, other memory-saving techniques have been implemented. For example, an edge-merging algorithm is available that writes sorted batches of edges to disk and processes them in order, rather than holding a large number of them in memory at once.

## 4 DATA AVAILABILITY

Babel outputs, encompassing compendia, synonym files, conflation mappings, and KGX exports, are hosted on RENCI infrastructure (https://stars.renci.org/var/babel/2025dec11-umls-level-0/), from where they can be reused in broader translational informatics applications – for instance, by loading them into a local SQLite database for rapid querying. Before publication, the most recent Babel release from these output files will be uploaded to UNC Dataverse and archived for long-term storage.

Identifiers can also be normalized using the publicly accessible Node Normalizer instance hosted at RENCI, available at https://nodenormalization-sri.renci.org/docs. Alternatively, a text search of all identifiers can be carried out using the Name Resolver at https://name-resolution-sri.renci.org/docs. Both services have a `/status` endpoint that can be used to look up the current version of Babel loaded at these instances. NCATS Translator maintains a production instance of these tools at https://nodenorm.transltr.io/ (Node Normalizer) and https://name-lookup.transltr.io/docs (Name Resolver) respectively. A Jupyter Notebook showing examples of Node Normalizer usage is available in its GitHub repository. You can contact other users of these services by using our mailing list at https://lists.renci.org/mailman3/lists/babel-tools.lists.renci.org/.

Graphs built using ORION for ROBOKOP can be downloaded, explored using a browser-based user interface, or queried programmatically at https://robokop.renci.org/.

## 5 Code Availability

The complete Babel software stack, including all Snakemake workflows, Python modules, Dockerfiles, and Helm configurations, is released under an MIT license and maintained in public repositories (https://github.com/NCATSTranslator/Babel). The Node Normalizer and the Name Resolver services are likewise open source, with deployment instructions and operational documentation available in their respective GitHub projects. Because Babel processes very large equivalence spaces, particularly for categories such as proteins, small molecules, and genes, its full execution requires substantial computational resources. Typical runs demand approximately 500 GB of RAM to support in-memory construction and validation of cliques.

The source code for Node Normalizer is openly available under an MIT license at https://github.com/NCATSTranslator/NodeNormalization. A Jupyter Notebook showing examples of Node Normalizer usage is available in its GitHub repository.

The source code for the Name Resolver is available under an MIT license at https://github.com/NCATSTranslator/NameResolution. A Jupyter Notebook showing examples of Name Resolver usage is available in its GitHub repository.

The entire ORION software stack is open source and is released under the MIT License in a public GitHub repository (https://github.com/RobokopU24/ORION). The GitHub repository includes Dockerfiles, Docker Compose configurations, and Helm charts for a variety of deployment options. GitHub Actions workflows are included for automatically running pytest tests and for publishing images to the GitHub Container Registry for each new release. A Jupyter Notebook showing examples of ORION usage is available in its GitHub repository.

## 6 Discussion

Interoperability, as defined for FAIR, is an attribute of a particular knowledge base on its own, rather than a pairwise attribute describing how source A can be used with source B. Instead, an interoperable data set A on its own is sufficiently well-described that a motivated user could connect it to other equally well-described datasets. This approach has been extremely important and influential in promoting the overall reuse of publicly funded biomedical data; however, producing integrated datasets from independently interoperable inputs continues to place a burden on downstream users, thereby reducing overall reuse.

Furthermore, the biomedical community is unlikely to fully standardize on a common set of entity identifiers or a single data model so that data sets are natively interoperable. Principled or arbitrary choices of data model or identifier schemes will always be a feature of independently produced data. Therefore, the community requires tools that fill the gap between interoperability in theory and interoperation in practice.

The Babel / ORION system is a pragmatic toolset for creating interoperable datasets. Fundamentally, this system is based on the recognition that even though all models are wrong and all abstractions are leaky, choices about models and identifiers must be made to create interoperation. Indeed, it is useful to share these choices with a larger community, even when they are imperfect, as they will always be. Babel outputs have already been used not just within the NCATS Biomedical Data Translator project that funded its development, but also by ROBOKOP (Bizon et al. 2019), the Dug semantic search engine (Waldrop et al. 2022), DrugMechDB (Gonzalez-Cavazos et al. 2023), the MeDIC project (DeLuca et al. 2025), the Every Cure ML/AI-Aided Therapeutic Repurposing In eXtended (MATRIX) project (in development), and the NCATS Ontology and Data Integration Network (ODIN) project (in development).

Recognition of a system's imperfections is not the same as resignation; we believe that the informatics community may find these tools useful as they currently exist, and that their continued use will drive improvements to the underlying biolink model and Babel's identifier mappings. Further, we anticipate adding flexibility to the underlying tools. In particular, conflation as a concept represents a form of flexibility in identifier clique formation that we plan to extend to other domains, such as the conflation of closely similar chemical structures or potentially orthologous genes. We also plan to allow users to configure Babel by allowing them to turn off pipelines or add new mappings as needed for specific applications. Similarly, flexibility in ORION will enable users to create specific, use case–driven integrations and datasets, moving toward the vision of truly interoperable and reusable biomedical data.

Finally, we welcome community input on improvements of the tools and resources described herein, namely, Babel, Node Normalizer, Name Resolver, and the ORION data integration and harmonization pipeline. All of these tools have been released under open source licenses. Any issues with running these pipelines or operating any of these services, or any contributions as pull requests, should be reported in the GitHub repositories listed under the "Code Availability" section. The Babel repository contains a "Contributors" document that provides specific instructions on steps you can take to make your contributions as easy to incorporate into the source code as possible. The Babel Tools mailing list can also be used to connect with other users of Babel, Node Normalizer and Name Resolver.

## Acknowledgements

We thank Jon Michael Beasley (AbbVie), Olawumi Olasunkanmi (RENCI), Hina Shah (RENCI), Shalki Shrivastava (LBNL), and Hong Yi (RENCI) for their contributions to the development and maintenance of the project repositories. Their efforts in software development, documentation, testing, and data curation were essential to advancing this work.

We also acknowledge the members of the NIH NCATS Biomedical Data Translator Consortium, whose discussions, issue reports, and conceptual feedback substantially informed the design and refinement of this framework. Their collaborative engagement helped to identify key integration challenges and prioritize features critical to interoperability across knowledge bases. The development of the Babel pipelines in particular required many people to plan, discuss, test, report issues and contribute ideas and tooling to this project. We would like to particularly thank Kent Shefcheck (Helix), Marian Mersmann, Eric Deutsch (Institute for Systems Biology), Colleen Xu (The Scripps Research Institute), Amy K Glen (Phenome Health), Andrew Su (The Scripps Research Institute), Mark D. Williams (NCATS), Richa Kanwar (RENCI), Kamileh Narsinh (Institute for Systems Biology), Stephen Ramsey (Oregon State University) and Gwênlyn Glusman (Institute for Systems Biology).

Beyond the Translator Consortium we are grateful to the broader research and open-source community, whose tools and resources provided critical foundations for the computational framework described herein.

## Funding

This research was supported by joint funding from the National Institute of Environmental Health Sciences (NIEHS) and the Office of Data Science Strategy within the National Institutes of Health under award **U24ES035214**, as well as by the National Center for Advancing Translational Sciences (NCATS) of the National Institutes of Health under award **OT2TR005712** (DOGSLED). The content is solely the responsibility of the authors and does not necessarily represent the official views of the National Institutes of Health.

# Supplemental Data

### Jupyter Notebooks

The included Jupyter Notebooks provide examples of Node Normalizer and Name Resolution usage.

- Node Normalizer: https://github.com/NCATSTranslator/NodeNormalization/blob/master/documentation/NodeNormalization.ipynb
- Name Resolver: https://github.com/NCATSTranslator/NameResolution/blob/master/documentation/NameResolution.ipynb

# Supplemental Tables

**Supplemental Table 1. Identifier Counts in Babel by CURIE Prefix and Biolink Type.** Number of identifiers in Babel 2025dec11, grouped by the CURIE prefix, which indicates the vocabulary that the identifier comes from. Some identifiers are known to incorrectly appear more than once in Babel, since each Babel pipeline independently sources identifiers for its files in parallel: this table therefore provides two counts for CURIEs: the total number of identifiers ("CURIE count") and the distinct number of identifiers ("Distinct CURIE count"). A breakdown by Biolink type is also provided.

| Prefix | CURIE count | Distinct CURIE count | Breakdown by Biolink type |
|---|---|---|---|
| **UniProtKB** | 199,651,241 | 199,651,241 | - biolink:Protein: 199,651,241 CURIEs |
| **PUBCHEM.COMPOUND** | 125,139,588 | 125,139,588 | - biolink:SmallMolecule: 111,547,805 CURIEs<br>- biolink:MolecularMixture: 11,716,398 CURIEs<br>- biolink:ChemicalEntity: 1,875,385 CURIEs |
| **INCHIKEY** | 115,976,745 | 115,976,745 | - biolink:SmallMolecule: 104,165,630 CURIEs<br>- biolink:MolecularMixture: 9,801,481 CURIEs<br>- biolink:ChemicalEntity: 2,009,633 CURIEs<br>- biolink:ChemicalMixture: 1 CURIEs |
| **NCBIGene** | 63,742,099 | 63,742,099 | - biolink:Gene: 63,742,099 CURIEs |
| **PMID** | 39,682,478 | 39,682,478 | - biolink:Publication: 39,682,478 CURIEs |
| **ENSEMBL** | 38,268,117 | 38,261,517 | - biolink:Protein: 21,279,080 CURIEs (21,272,480 distinct)<br>- biolink:Gene: 17,002,237 CURIEs (16,995,637 distinct) |
| **doi** | 30,758,070 | 30,758,070 | - biolink:Publication: 30,758,070 CURIEs |
| **PMC** | 10,658,639 | 10,658,639 | - biolink:Publication: 10,658,639 CURIEs |
| **CAS** | 4,127,675 | 4,127,675 | - biolink:SmallMolecule: 3,813,016 CURIEs<br>- biolink:MolecularMixture: 307,963 CURIEs<br>- biolink:ChemicalEntity: 6,689 CURIEs<br>- biolink:ComplexMolecularMixture: 4 CURIEs<br>- biolink:Drug: 3 CURIEs |
| **NCBITaxon** | 2,709,586 | 2,709,586 | - biolink:OrganismTaxon: 2,709,586 CURIEs |
| **CHEMBL.COMPOUND** | 2,480,347 | 2,480,347 | - biolink:SmallMolecule: 2,233,709 CURIEs<br>- biolink:ChemicalEntity: 135,913 CURIEs<br>- biolink:MolecularMixture: 110,725 CURIEs |
| **UMLS** | 2,459,955 | 2,459,378 | - biolink:OrganismTaxon: 769,724 CURIEs<br>- biolink:Procedure: 244,699 CURIEs<br>- biolink:Drug: 225,570 CURIEs<br>- biolink:ChemicalEntity: 200,799 CURIEs (200,798 distinct)<br>- biolink:Protein: 160,925 CURIEs<br>- biolink:PhenotypicFeature: 120,794 CURIEs<br>- biolink:AnatomicalEntity: 119,934 CURIEs (119,933 distinct)<br>- biolink:ClinicalAttribute: 109,956 CURIEs<br>- biolink:Disease: 107,657 CURIEs<br>- biolink:SmallMolecule: 72,846 CURIEs<br>- biolink:Gene: 46,912 CURIEs<br>- biolink:Publication: 45,262 CURIEs<br>- biolink:InformationContentEntity: 44,812 CURIEs<br>- biolink:GenomicEntity: 41,922 CURIEs<br>- biolink:BiologicalProcess: 39,927 CURIEs<br>- biolink:MolecularActivity: 28,396 CURIEs<br>- biolink:Device: 21,460 CURIEs<br>- biolink:Activity: 9,860 CURIEs<br>- biolink:CellularComponent: 8,141 CURIEs<br>- biolink:MolecularMixture: 6,326 CURIEs<br>- biolink:Cell: 6,027 CURIEs (6,026 distinct)<br>- biolink:GeographicLocation: 4,262 CURIEs<br>- biolink:PopulationOfIndividualOrganisms: 4,106 CURIEs<br>- biolink:Phenomenon: 3,940 CURIEs<br>- biolink:Cohort: 3,580 CURIEs<br>- biolink:Behavior: 2,870 CURIEs<br>- biolink:PhysicalEntity: 2,845 CURIEs<br>- biolink:Agent: 2,704 CURIEs<br>- biolink:GrossAnatomicalStructure: 2,380 CURIEs<br>- biolink:NucleicAcidEntity: 454 CURIEs<br>- biolink:Polypeptide: 376 CURIEs<br>- biolink:Event: 135 CURIEs<br>- biolink:PhysiologicalProcess: 111 CURIEs<br>- biolink:BiologicalEntity: 109 CURIEs<br>- biolink:ComplexMolecularMixture: 87 CURIEs<br>- biolink:Human: 46 CURIEs<br>- biolink:DiseaseOrPhenotypicFeature: 27 CURIEs<br>- biolink:NamedThing: 23 CURIEs<br>- biolink:MolecularEntity: 13 CURIEs<br>- biolink:ChemicalMixture: 6 CURIEs |
| **MESH** | 446,681 | 346,827 | - biolink:ChemicalEntity: 170,767 CURIEs (170,743 distinct)<br>- biolink:Protein: 105,470 CURIEs<br>- biolink:SmallMolecule: 78,742 CURIEs<br>- biolink:OrganismTaxon: 70,718 CURIEs (70,653 distinct)<br>- biolink:Disease: 11,267 CURIEs<br>- biolink:MolecularMixture: 6,608 CURIEs<br>- biolink:PhenotypicFeature: 982 CURIEs<br>- biolink:GrossAnatomicalStructure: 779 CURIEs<br>- biolink:AnatomicalEntity: 644 CURIEs (607 distinct)<br>- biolink:Cell: 340 CURIEs (331 distinct)<br>- biolink:CellularComponent: 284 CURIEs (280 distinct)<br>- biolink:ComplexMolecularMixture: 183 CURIEs (182 distinct)<br>- biolink:Drug: 29 CURIEs<br>- biolink:ChemicalMixture: 8 CURIEs |
| **RGD** | 317,397 | 317,397 | - biolink:Gene: 317,397 CURIEs |
| **PR** | 253,470 | 253,470 | - biolink:Protein: 253,470 CURIEs |
| **HMDB** | 217,920 | 217,920 | - biolink:SmallMolecule: 217,469 CURIEs<br>- biolink:MolecularMixture: 362 CURIEs<br>- biolink:ChemicalEntity: 89 CURIEs |
| **CHEBI** | 202,960 | 202,960 | - biolink:SmallMolecule: 172,720 CURIEs<br>- biolink:ChemicalEntity: 26,121 CURIEs<br>- biolink:MolecularMixture: 3,582 CURIEs<br>- biolink:ChemicalMixture: 537 CURIEs |
| **UNII** | 139,647 | 139,647 | - biolink:SmallMolecule: 67,158 CURIEs<br>- biolink:ChemicalEntity: 55,679 CURIEs<br>- biolink:MolecularMixture: 16,809 CURIEs<br>- biolink:ChemicalMixture: 1 CURIEs |
| **RXCUI** | 124,874 | 124,874 | - biolink:Drug: 102,786 CURIEs<br>- biolink:ChemicalEntity: 13,214 CURIEs<br>- biolink:MolecularMixture: 4,529 CURIEs<br>- biolink:SmallMolecule: 4,345 CURIEs |
| **REACT** | 112,754 | 112,754 | - biolink:MolecularActivity: 89,878 CURIEs<br>- biolink:Pathway: 22,358 CURIEs<br>- biolink:BiologicalProcess: 518 CURIEs |
| **FMA** | 98,633 | 98,633 | - biolink:AnatomicalEntity: 98,633 CURIEs |
| **MGI** | 94,277 | 94,277 | - biolink:Gene: 94,277 CURIEs |
| **RHEA** | 71,132 | 71,132 | - biolink:MolecularActivity: 71,132 CURIEs |
| **NCIT** | 59,588 | 59,531 | - biolink:Disease: 27,359 CURIEs<br>- biolink:PhenotypicFeature: 21,890 CURIEs<br>- biolink:AnatomicalEntity: 4,224 CURIEs<br>- biolink:Cell: 2,577 CURIEs<br>- biolink:GrossAnatomicalStructure: 2,081 CURIEs<br>- biolink:CellularComponent: 1,457 CURIEs |
| **WormBase** | 48,778 | 48,778 | - biolink:Gene: 48,778 CURIEs |
| **HGNC** | 44,741 | 44,741 | - biolink:Gene: 44,741 CURIEs |
| **GO** | 39,354 | 39,354 | - biolink:BiologicalProcess: 24,568 CURIEs<br>- biolink:MolecularActivity: 10,143 CURIEs<br>- biolink:CellularComponent: 4,058 CURIEs<br>- biolink:Pathway: 585 CURIEs |
| **CLO** | 38,810 | 38,810 | - biolink:CellLine: 38,810 CURIEs |
| **ZFIN** | 37,990 | 37,990 | - biolink:Gene: 37,990 CURIEs |
| **FB** | 30,256 | 30,256 | - biolink:Gene: 30,256 CURIEs |
| **SMPDB** | 30,248 | 30,248 | - biolink:Pathway: 30,248 CURIEs |
| **OMIM** | 28,018 | 27,969 | - biolink:Gene: 17,759 CURIEs<br>- biolink:Disease: 10,259 CURIEs |
| **MONDO** | 26,333 | 26,333 | - biolink:Disease: 26,333 CURIEs |
| **PANTHER.FAMILY** | 26,140 | 26,140 | - biolink:GeneFamily: 26,140 CURIEs |
| **medgen** | 21,139 | 21,138 | - biolink:Disease: 21,139 CURIEs (21,138 distinct) |
| **HP** | 19,076 | 19,076 | - biolink:PhenotypicFeature: 16,707 CURIEs<br>- biolink:Disease: 2,369 CURIEs |
| **DRUGBANK** | 17,678 | 16,547 | - biolink:SmallMolecule: 11,571 CURIEs<br>- biolink:ChemicalEntity: 3,147 CURIEs<br>- biolink:Protein: 2,447 CURIEs<br>- biolink:MolecularMixture: 509 CURIEs<br>- biolink:ComplexMolecularMixture: 3 CURIEs<br>- biolink:Drug: 1 CURIEs |
| **KEGG.COMPOUND** | 16,066 | 16,066 | - biolink:SmallMolecule: 14,039 CURIEs<br>- biolink:ChemicalEntity: 1,502 CURIEs<br>- biolink:MolecularMixture: 468 CURIEs<br>- biolink:ChemicalMixture: 57 CURIEs |
| **UBERON** | 14,608 | 14,608 | - biolink:GrossAnatomicalStructure: 10,507 CURIEs<br>- biolink:AnatomicalEntity: 4,090 CURIEs<br>- biolink:CellularComponent: 8 CURIEs<br>- biolink:Cell: 3 CURIEs |
| **dictyBase** | 13,892 | 13,892 | - biolink:Gene: 13,892 CURIEs |
| **GTOPDB** | 13,510 | 13,510 | - biolink:SmallMolecule: 13,231 CURIEs<br>- biolink:ChemicalEntity: 211 CURIEs<br>- biolink:MolecularMixture: 68 CURIEs |
| **DOID** | 12,012 | 12,012 | - biolink:Disease: 12,012 CURIEs |
| **SNOMEDCT** | 11,953 | 11,953 | - biolink:Disease: 9,828 CURIEs<br>- biolink:PhenotypicFeature: 2,125 CURIEs |
| **EFO** | 11,586 | 11,586 | - biolink:PhenotypicFeature: 7,859 CURIEs<br>- biolink:Disease: 3,727 CURIEs |
| **orphanet** | 11,048 | 11,048 | - biolink:Disease: 11,048 CURIEs |
| **EC** | 8,813 | 8,813 | - biolink:MolecularActivity: 8,813 CURIEs |
| **SGD** | 7,167 | 7,167 | - biolink:Gene: 7,167 CURIEs |
| **DrugCentral** | 4,995 | 4,995 | - biolink:SmallMolecule: 4,302 CURIEs<br>- biolink:ChemicalEntity: 388 CURIEs<br>- biolink:MolecularMixture: 305 CURIEs |
| **CL** | 3,160 | 3,160 | - biolink:Cell: 3,159 CURIEs<br>- biolink:CellularComponent: 1 CURIEs |
| **ICD10** | 2,520 | 2,520 | - biolink:Disease: 2,520 CURIEs |
| **ICD9** | 2,233 | 2,233 | - biolink:Disease: 2,233 CURIEs |
| **HGNC.FAMILY** | 1,999 | 1,999 | - biolink:GeneFamily: 1,999 CURIEs |
| **EMAPA** | 966 | 966 | - biolink:AnatomicalEntity: 966 CURIEs |
| **MEDDRA** | 732 | 732 | - biolink:Disease: 534 CURIEs<br>- biolink:PhenotypicFeature: 198 CURIEs |
| **ComplexPortal** | 631 | 631 | - biolink:MacromolecularComplex: 631 CURIEs |
| **ZFA** | 606 | 606 | - biolink:AnatomicalEntity: 606 CURIEs |
| **OMIM.PS** | 572 | 572 | - biolink:Disease: 572 CURIEs |
| **PANTHER.PATHWAY** | 175 | 175 | - biolink:Pathway: 175 CURIEs |
| **FBbt** | 117 | 117 | - biolink:AnatomicalEntity: 117 CURIEs |
| **KEGG.DISEASE** | 40 | 40 | - biolink:Disease: 40 CURIEs |
| **MP** | 33 | 33 | - biolink:PhenotypicFeature: 30 CURIEs<br>- biolink:Disease: 3 CURIEs |
| **TCDB** | 21 | 21 | - biolink:MolecularActivity: 21 CURIEs |
| **WBbt** | 18 | 18 | - biolink:AnatomicalEntity: 18 CURIEs |
| **KEGG.REACTION** | 9 | 9 | - biolink:MolecularActivity: 9 CURIEs |
| **icd11** | 5 | 5 | - biolink:Disease: 5 CURIEs |

**Supplemental Table 2. The fourteen output pipelines that form part of Babel.** Each pipeline produces identifiers of one or more Biolink files. The number of cliques and the number of identifiers and distinct synonyms are included, as well as the CURIE prefixes (ordered from the prefix with the largest number of identifiers to the smallest).

| Pipeline | Biolink Types | Number of CURIEs | Number of distinct CURIEs | Clique leader prefixes | CURIE prefixes |
|---|---|---|---|---|---|
| **Anatomy** Anatomical entities at all scales, from brains to endothelium to pancreatic beta cells | AnatomicalEntity | 227,885 | 227,885 | MESH, NCIT, UBERON, UMLS | FMA, ZFA, UMLS, MESH, WBbt, EMAPA, UBERON, NCIT, FBbt |
| | Cell | 12,088 | 12,088 | CL, MESH, NCIT, UMLS | CL, MESH, UMLS, UBERON, NCIT |
| | CellularComponent | 13,945 | 13,945 | GO, MESH, NCIT, UMLS | CL, MESH, UMLS, GO, UBERON, NCIT |
| | GrossAnatomicalStructure | 15,747 | 15,747 | UBERON | NCIT, MESH, UMLS, UBERON |
| **CellLine** Cell lines from different species | CellLine | 38,810 | 38,810 | CLO | CLO |
| **Chemicals** All kinds of chemicals, including drugs, small molecules, molecular mixtures, and so on | MolecularMixture | 21,976,133 | 21,976,133 | CHEBI, CHEMBL.COMPOUND, HMDB, PUBCHEM.COMPOUND, RXCUI, UNII | CHEMBL.COMPOUND, KEGG.COMPOUND, HMDB, GTOPDB, UNII, MESH, UMLS, DrugCentral, PUBCHEM.COMPOUND, CHEBI, DRUGBANK, RXCUI, CAS, INCHIKEY |
| | SmallMolecule | 222,416,583 | 222,416,583 | CHEBI, CHEMBL.COMPOUND, DrugCentral, GTOPDB, HMDB, PUBCHEM.COMPOUND, UNII | CHEMBL.COMPOUND, KEGG.COMPOUND, INCHIKEY, GTOPDB, UNII, MESH, UMLS, DrugCentral, PUBCHEM.COMPOUND, CHEBI, DRUGBANK, RXCUI, CAS, HMDB |
| | Polypeptide | 163 | 163 | UMLS | UMLS |
| | ComplexMolecularMixture | 275 | 275 | DRUGBANK, MESH | MESH, UMLS, DRUGBANK, CAS |
| | ChemicalEntity | 4,488,540 | 4,488,540 | CHEBI, CHEMBL.COMPOUND, DRUGBANK, DrugCentral, KEGG.COMPOUND, MESH, PUBCHEM.COMPOUND, RXCUI, UMLS, UNII | CHEMBL.COMPOUND, KEGG.COMPOUND, HMDB, GTOPDB, MESH, UMLS, UNII, DrugCentral, PUBCHEM.COMPOUND, CHEBI, DRUGBANK, RXCUI, CAS, INCHIKEY |
| | ChemicalMixture | 610 | 610 | CHEBI | KEGG.COMPOUND, MESH, UMLS, UNII, CHEBI, INCHIKEY |
| | Drug | 314,727 | 314,727 | RXCUI, UMLS | MESH, UMLS, DRUGBANK, RXCUI, CAS |
| **Diseases and Phenotypes** | Disease | 248,905 | 248,904 | DOID, EFO, MESH, MONDO, NCIT, OMIM, UMLS, orphanet | MP, KEGG.DISEASE, SNOMEDCT, MEDDRA, icd11, ICD9, MESH, UMLS, OMIM.PS, MONDO, medgen, ICD10, orphanet, EFO, NCIT, OMIM, HP, DOID |
| | PhenotypicFeature | 170,585 | 170,585 | EFO, HP, MESH, NCIT, UMLS | MP, SNOMEDCT, MEDDRA, MESH, UMLS, EFO, NCIT, HP |
| **DrugChemical**<br>Conflation of drugs with their active ingredients as chemicals | 9,098 sets of conflations, with between 2 and 729 cliques represented by their preferred CURIEs. There are a total of 106,199 preferred CURIEs with no duplicates. | | | | |
| **Gene**<br>Genes from all species | Gene | 81,396,905 | 81,396,905 | ENSEMBL, FB, HGNC, MGI, NCBIGene, OMIM, RGD, SGD, UMLS, WormBase, ZFIN | dictyBase, NCBIGene, HGNC, ENSEMBL, MGI, UMLS, SGD, RGD, ZFIN, WormBase, FB, OMIM |
| **GeneFamily**<br>Families of genes | GeneFamily | 28,139 | 28,139 | HGNC.FAMILY, PANTHER.FAMILY | HGNC.FAMILY, PANTHER.FAMILY |
| **GeneProtein**<br>Conflation of genes with the proteins they code for. | 15,051,450 sets of conflations, with between 2 and 609 cliques represented by their preferred CURIEs. There are a total of 32,781,254 preferred CURIEs with no duplicates. | | | | |
| **Leftover UMLS**<br>A special pipeline that adds every UMLS concept not already added elsewhere in Babel | umls | 569,764 | 569,764 | UMLS | UMLS |
| **Macromolecular Complex** | MacromolecularComplex | 631 | 631 | ComplexPortal | ComplexPortal |
| **ProcessActivityPathway**<br>Biological processes, activities and pathways | Pathway | 53,366 | 53,366 | GO, PANTHER.PATHWAY, REACT, SMPDB | SMPDB, REACT, PANTHER.PATHWAY, GO |
| | BiologicalProcess | 65,013 | 65,013 | GO, REACT, UMLS | REACT, UMLS, GO |
| | MolecularActivity | 208,097 | 208,097 | EC, GO, REACT, RHEA, UMLS | TCDB, UMLS, GO, RHEA, EC, KEGG.REACTION, REACT |
| **Protein**<br>Proteins from all species | Protein | 221,445,861 | 221,445,861 | ENSEMBL, PR, UMLS, UniProtKB | PR, ENSEMBL, UniProtKB, MESH, UMLS, DRUGBANK |
| **Publications**<br>All publications from PubMed | Publication | 81,099,187 | 81,099,187 | PMID | PMID, PMC, doi |
| **Taxon**<br>Taxonomic entities, including species, genera, families, and so on from the NCBI Taxonomy | OrganismTaxon | 3,549,962 | 3,549,962 | MESH, NCBITaxon, UMLS | MESH, UMLS, NCBITaxon |

**Supplemental Table 3. A list of mapping sources used by Babel in combining identifiers from different sources.** In some cases (e.g. the UBERON ontology), we include every cross-reference from the mapping source, while in others (e.g. Wikidata) we only extract a subset of the mappings, sometimes with additional filtering. In this case, the "Number of mappings" column lists the number of cross-references between pairs of CURIE prefixes (e.g. 586 cross-references between cell ontology identifiers starting with "CL:" and UMLS identifiers starting with "UMLS:"). In some cases (e.g. "HGNC Gene Family labels"), all identifiers are included without any cross-mappings.

| Biolink Types | Mapping Source | Number of mappings |
|---|---|---|
| AnatomicalEntity<br>Cell<br>CellularComponent<br>GrossAnatomicalStructure | UBERON ontology | 38,910 cross-references from UBERON:0001062 "anatomical entity" and its subclasses. |
| | GO ontology | 305 cross-references from GO:0005575 "cellular_component" and its subclasses. |
| | Wikidata | *Query Wikidata for entities that have both a wd:P7963 ("Cell Ontology ID") and wd:P2892 ("UMLS CUI"), and then generate cross-references between them.*<br>CL, UMLS: 596 |
| | UMLS | FMA, UMLS: 103,205<br>GO, UMLS: 5,412<br>MESH, UMLS: 1,794<br>NCIT, UMLS: 8,902 |
| BiologicalProcess<br>MolecularActivity<br>Pathway | RHEA | EC, RHEA: 7,858 |
| | GO | 26,512 cross-references from GO:0007165, GO:0008150 and GO:0003674 and its subclasses. |
| | UMLS | GO, UMLS: 53,274 |
| CellLine | Cell Line Ontology | 38,810 identifiers |
| Gene | MIM2Gene MEDGEN | NCBIGene, OMIM: 17,461 |
| | NCBIGene gene2ensembl.gz | ENSEMBL, NCBIGene: 13,467,445 |
| | NCBIGene gene_info.gz | FB, NCBIGene: 17,872<br>HGNC, NCBIGene: 43,795<br>MGI, NCBIGene: 58,200<br>NCBIGene, RGD: 303,125<br>NCBIGene, SGD: 6,471<br>NCBIGene, WormBase: 46,890<br>NCBIGene, ZFIN: 27,274<br>NCBIGene, dictyBase: 13,892 |
| | UMLS-UniProtKB mappings | NCBIGene, UMLS: 23,036 |
| | UMLS | HGNC, UMLS: 44,066 |
| GeneFamily | PANTHER Sequence Classification: Human | All identifiers |
| | HGNC Gene Family labels | All identifiers |
| MacromolecularComplex | ComplexPortal for organism 559292 (Saccharomyces cerevisiae) | 631 identifiers |
| OrganismTaxon | MeSH | MESH, NCBITaxon: 70,106 |
| | UMLS | MESH, UMLS: 69,247 |
| Publication | PubMed Baseline and updates | All identifiers |
| | PMC/PMID mappings from PMC | PMC, PMID: 12,593,920<br>PMID, doi: 35,216,885 |
| ChemicalEntity<br>ChemicalMixture<br>ComplexMolecularMixture<br>Drug<br>MolecularMixture<br>Polypeptide<br>SmallMolecule | DrugCentral | CHEBI, DrugCentral: 4,302<br>CHEMBL.COMPOUND, DrugCentral: 7,129<br>DRUGBANK, DrugCentral: 4,368<br>DrugCentral, GTOPDB: 2,106<br>DrugCentral, MESH: 2,014<br>DrugCentral, PUBCHEM.COMPOUND: 5,013<br>DrugCentral, UMLS: 5,244<br>DrugCentral, UNII: 5,185 |
| | MeSH | CAS, MESH: 48,440 |
| | MeSH | MESH, UNII: 18,077 |
| | WikiData | CHEBI, MESH: 6,208 |
| | ChEBI | CHEBI, KEGG.COMPOUND: 15,638 |
| | GtoPDB | GTOPDB, INCHIKEY: 11,177 |
| | PubChem CAS | CAS, PUBCHEM.COMPOUND: 4,220,873 |
| | PubChem MeSH | MESH, PUBCHEM.COMPOUND: 77,747 |
| | UMLS | DRUGBANK, UMLS: 8,003<br>MESH, UMLS: 148,791<br>RXCUI, UMLS: 116,638 |
| | UMLS | DRUGBANK, RXCUI: 5,212 |
| Disease<br>PhenotypicFeature | EFO ontology | 28,559 cross-references |
| | DOID | 38,566 cross-references |
| | HP ontology | 17,962 cross-references |
| | MONDO ontology | 107,021 cross-references from the descendants of MONDO:0000001 and MONDO:0042489, including ORPHANET prefixes |
| | UMLS | HP, UMLS: 20,664<br>MESH, UMLS: 12,082<br>NCIT, UMLS: 48,081<br>OMIM, UMLS: 11,616 |
| | Manual concords | DOID, MONDO: 3<br>EFO, MONDO: 1<br>EFO, UMLS: 7<br>HP, NCIT: 1<br>HP, UMLS: 1<br>MONDO, UMLS: 63<br>UMLS, UMLS: 8<br>HP, MONDO: 2<br>MESH, MONDO: 1 |
| DrugChemical conflation | Manual concords | 32 cross-references |
| | RxNorm mappings from PubChem | 4,246 cross-references |
| | RxNorm mappings | 310,002 cross-references (although not all of these are used) |
| Protein | NCIt-SwissProt_Mapping | NCIT, UniProtKB: 6,277 |
| | PR | 205,386 cross-references from subclasses and xrefs of PR:000000001 |
| | UniProtKB idmapping.dat | ENSEMBL, ENSEMBL: 9,673,501<br>ENSEMBL, UniProtKB: 9,731,073 |
| | UMLS GeneProtein | UMLS, UniProtKB: 23,036 |
| | UMLS | NCIT, UMLS: 15,478<br>DRUGBANK, UMLS: 2,447<br>MESH, UMLS: 105,470 |

**Supplemental Table 4. Knowledge Sources with Available ORION Parsers.** This table lists upstream biomedical knowledge sources for which dedicated parsers have been implemented in ORION. Each parser retrieves versioned data releases from the source and transforms them into a standardized intermediate representation suitable for downstream normalization and integration. Inclusion indicates parser availability and does not necessarily imply inclusion in a specific integrated knowledge graph release.

| Available Parsers | Description |
|---|---|
| **BindingDB**<br>https://www.bindingdb.org/ | A public, web-accessible database of measured binding affinities, focusing chiefly on the interactions of proteins considered to be candidate drug-targets with ligands that are small, drug-like molecules. |
| **CAM KP**<br>https://github.com/ExposuresProvider/cam-kp-api | CAMs (Causal Activity Models) are small knowledge graphs built using the Web Ontology Language (OWL). The CAM database combines many CAM graphs along with a large merged bio-ontology containing the full vocabulary of concepts referenced within the individual CAMs. Each CAM describes an instantiation of some of those concepts in a particular context, modeling the interactions between those instances as an interlinked representation of a complex biological or environmental process. |
| **ChEBI**<br>https://www.ebi.ac.uk/chebi/ | Chemical Entities of Biological Interest (ChEBI) is a freely available dictionary of molecular entities focused on 'small' chemical compounds. |
| **Clinical Trials KP**<br>https://aact.ctti-clinicaltrials.org/ | The Clinical Trials KP, created and maintained by the Multiomics Provider, provides information on Clinical Trials, ultimately derived from researcher submissions to clinicaltrials.gov, via the Aggregate Analysis of Clinical Trials (AACT) database). Information on select trials includes the NCT Identifier of the trial, interventions used, diseases/conditions relevant to the trial, adverse events, etc. |
| **COHD**<br>https://cohd.io/ | Columbia Open Health Data (COHD) provides access to counts and frequencies (i.e., prevalence in electronic health records) of conditions, procedures, drug exposures, and patient demographics among inpatients and outpatients, and the co-occurrence frequencies between them. The data are derived from the Observational Medical Outcomes Partnership (OMOP) database at Columbia University Irving Medical Center. |
| **DrugCentral**<br>https://drugcentral.org/ | DrugCentral provides open-source data on active ingredients, chemical entities, pharmaceutical products, drug mode of action, indications, and pharmacologic action for approved drugs, derived from information provided by the US Food & Drug Administration, the European Medicines Agency, and the Pharmaceutical and Medical Devices Agency. Caveat: DrugCentral provides limited information on discontinued and drugs approved outside of the US, but users should be aware that that information has not been validated. |
| **DrugMechDB**<br>https://sulab.github.io/DrugMechDB/ | A database of paths that represent the mechanism of action from a drug to a disease in an indication. |
| **GTEx**<br>https://www.gtexportal.org/ | The Genotype-Tissue Expression (GTEx) portal provides open access to data on tissue-specific gene expression and regulation, derived from molecular assays (e.g., WGS, WES, RNA-Seq) on 54 non-diseased tissue sites across nearly 1000 individuals. |
| **GWAS Catalog**<br>https://www.ebi.ac.uk/gwas/ | The Genome-Wide Association Studies (GWAS) Catalog provides a curated, searchable, visualisable, and openly available database of single nucleotide polymorphism (SNP)-trait associations, derived from all GWAS publications identified by curators, who then extract the reported trait, significant SNP-trait associations, and sample metadata. |
| **Alliance of Genome Resources**<br>https://www.alliancegenome.org/ | A consortium of 7 model organism databases (MODs) and the Gene Ontology (GO) Consortium whose goal is to provide an integrated view of their data to all biologists, clinicians and other interested parties. |
| **HGNC**<br>https://www.genenames.org/about/ | The HUGO Gene Nomenclature Committee (HGNC) database provides open access to HGNC-approved unique symbols and names for human genes, gene groups, and associated resources, including links to genomic, proteomic and phenotypic information. |
| **HMDB**<br>https://hmdb.ca/ | The Human Metabolome Database (HMDB) is an openly accessible database that contains detailed information about small molecule metabolites found in the human body, with links between chemical data, clinical data, and molecular biology/biochemistry data, including protein sequences (such as enzymes and transporters). |
| **Hetio**<br>https://het.io/about/ | Hetionet is an open-source biomedical heterogeneous information network (hetnet) or graph-based resource describing relationships uncovered by millions of biomedical research studies over the past fifty years. |
| **Gene Ontologies Annotations**<br>https://www.ebi.ac.uk/GOA/ | The Gene Ontology (GO) Consortium's Human Gene Ontologies Annotations (Human GOA) resource provides open access to curated assignment of GO terms to proteins in the UniProt KnowledgeBase (UniProtKB), RNA molecules from RNACentral, and protein complexes from the Complex Portal. |
| **IntAct**<br>https://www.ebi.ac.uk/intact/ | The IntAct Molecular Interaction Database provides open access to molecular interactions data derived from literature curation or direct user submission. |
| **KinAce**<br>https://kinace.kinametrix.com/ | The KinAce web portal aggregates and visualizes the network of interactions between protein-kinases and their substrates in the human genome. |
| **Monarch KG**<br>https://monarchinitiative.org/ | The Monarch knowledge graph is a reference implementation of the Biolink Model specification. It contains data from the Monarch Initiative aggregated database, as well as several OBO ontologies. These data sources provide links between diseases, phenotypes, genomic features across human and non-human organisms. |
| **PANTHER**<br>http://pantherdb.org/ | The Protein ANalysis THrough Evolutionary Relationships (PANTHER) classification system provides an openly available annotation library of gene family phylogenetic trees, with persistent identifiers attached to all nodes in the trees and annotation of each protein member |
| | of the family by its family and protein class, subfamily, orthologs, paralogs, GO Phylogenetic Annotation Project function and Reactome pathways. |
| **PHAROS**<br>https://pharos.nih.gov/ | Pharos is the openly accessible user interface to the Illuminating the Druggable Genome (IDG) program's Knowledge Management Center (KMC), which aims to develop a comprehensive, integrated knowledge-base for the Druggable Genome (DG) to illuminate the uncharacterized and/or poorly annotated portion of the DG, focusing on three of the most commonly drug-targeted protein families: G-protein-coupled receptors; ion channels; and kinases. |
| **Reactome**<br>https://reactome.org/ | Reactome is a free, open-source, curated and peer-reviewed pathway database. |
| **STRING-DB**<br>https://string-db.org | The Search Tool for the Retrieval of Interacting Genes/Proteins (STRING) database provides information on known and predicted protein-protein interactions (both direct and indirect) derived from genomic context predictions, high-throughput laboratory experiments, conserved co-expression, automated text mining, and aggregated knowledge from primary data sources. |
| **Text Mining Provider KG**<br>https://github.com/NCATSTranslator/Translator-All/wiki/Text-Mining-Provider | The Text Mining Provider KG contains subject-predicate-object assertions derived from the application of natural language processing (NLP) algorithms to the PubMedCentral Open Access collection of publications plus additional titles and abstracts from PubMed. |
| **Ubergraph**<br>https://github.com/INCATools/ubergraph | Ubergraph is an open-source graph database containing integrated ontologies, including GO, CHEBI, HPO, and Uberon's anatomical ontology. |
| **ViralProteome**<br>https://www.ebi.ac.uk/GOA/proteomes | The Gene Ontology (GO) Consortium's Viral Proteome resource provides open access to curated assignment of GO terms to proteins and proteome relationships derived from the UniProt KnowledgeBase for all NCBI Taxa considered viruses. |